\documentclass[reprint, twocolumn,showpacs,superscriptaddress,preprintnumbers,amsmath,amssymb,aps,pre]{revtex4-1}

\usepackage{graphics,amssymb,amsmath,epsfig,color}
\usepackage{graphicx}
\usepackage{dcolumn}
\usepackage{bm}
\usepackage{color}

\begin{document}


\title{Polymer Glass-Formation in Variable Dimension}
\thanks{Contribution of the National Institute of Standards and Technology - Work not subject to copyright in the United States.}

\author{Wen-Sheng Xu}
\email{wsxu@uchicago.edu}
\affiliation{James Franck Institute, The University of Chicago, Chicago, Illinois 60637, USA}

\author{Jack F. Douglas}
\email{jack.douglas@nist.gov}
\affiliation{Materials Science and Engineering Division, National Institute of Standards and Technology, Gaithersburg, Maryland 20899, USA}

\author{Karl F. Freed}
\email{freed@uchicago.edu}
\affiliation{James Franck Institute, The University of Chicago, Chicago, Illinois 60637, USA}
\affiliation{Department of Chemistry, The University of Chicago, Chicago, Illinois 60637, USA}

\date{\today}

\begin{abstract}
We explore the nature of glass-formation in variable spatial dimensionality ($d$) based on the generalized entropy theory, a synthesis of the Adam-Gibbs model with direct computation of the configurational entropy of polymer fluids using an established statistical mechanical model. We find that structural relaxation in the fluid state asymptotically becomes Arrhenius in the $d\rightarrow\infty$ limit and that the fluid transforms upon sufficient cooling above a critical dimension near $d=8$ into a dense amorphous state with a finite positive residual configurational entropy. Direct computations of the isothermal compressibility and thermal expansion coefficient, taken to be physical measures of packing frustration, demonstrate that these fluid properties strongly correlate with the fragility of glass-formation.
\end{abstract}



\maketitle

Growing evidence indicates that molecular packing plays an essential role in both the physics of crystallization~\cite{JCP_88_1177, JCP_136_054106, JCP_137_074106, PRE_65_016108, PRE_80_061110} and glass-formation~\cite{PRE_74_041127, JCP_138_12A548, PNAS_106_15171, JCP_139_164502, PRL_109_095705, PRL_107_185702, RMP_82_789, PRL_104_255704, PRE_81_041502, PRE_82_051126, JPCM_19_205101, EPJE_34_102}, prompting a consideration of variable spatial dimensionality ($d$) as a conceptual probe of these solidification processes. In particular,  the ``decorrelation principle'', recently proposed by Torquato and Stillinger~\cite{PRE_73_031106, PRE_74_061308}, states that unconstrained correlations in hard hyperspheres diminish with increasing $d$ and vanish in the $d\rightarrow \infty$ limit. Likewise, recent simulations~\cite{JCP_138_12A548, PNAS_106_15171, JCP_139_164502} have indicated that certain often cited aspects of glass-formation, such as the dynamic heterogeneity and the decoupling between structural relaxation and diffusion, become diminished at elevated $d$. While prior studies~\cite{PRL_107_185702, RMP_82_789, PRL_104_255704, PRE_81_041502, PRE_82_051126} have focused mainly on how critical packing fractions associated with glass-formation in hard hyperspheres vary with $d$, little is known concerning the $d$ dependence of the characteristic temperatures and fragility of glass-formation in molecular fluids. No prior work has considered complex fluids, such as polymeric liquids, where many molecular parameters may be tuned to control the nature of glass-formation.
        
In this Letter, we consider the glass-formation of a model polymer glass-forming (GF) liquid in $d$ dimensions. Our approach is based on the generalized entropy theory (GET)~\cite{ACP_137_125}, which is a combination of the lattice cluster theory (LCT)~\cite{ACP_103_335} and the Adam-Gibbs (AG) theory~\cite{JCP_43_139} linking the configurational entropy $s_c$ (i.e., the entropy devoid of the vibrational component) to the structural relaxation time $\tau_{\alpha}$. The LCT employs a $d$-dimensional hypercubic lattice model, whose use facilitates the development of an expansion about the mean-field limit of infinite $d$. The LCT thus naturally provides a framework for considering the $d$ dependence of polymer thermodynamic properties, thereby making the GET suitable for exploring glass-formation in $d$ dimensions.

As a primary result, we find that structural relaxation in the fluid state asymptotically becomes Arrhenius in the $d\rightarrow \infty$ limit~\footnote{The LCT free energy contains a term $\ln(d)$~\cite{ACP_103_335}, which complicates the discussion of the $d\rightarrow \infty$ limit. By $d\rightarrow \infty$, we mean an arbitrarily large but finite $d$.}, a limiting behavior that is a dynamical counterpart of the decorrelation principle of Torquato and Stillinger~\cite{PRE_73_031106, PRE_74_061308}. Moreover, our theory indicates that the fluid transforms upon cooling above a critical dimension near $d=8$ into a dense amorphous state with a finite positive residual configurational entropy. We show that the ``kinetic fragility'', determined from the temperature ($T$) dependence of $\tau_{\alpha}$, decreases with $d$, while the ``thermodynamic fragility'', defined as the relative rate of change of $s_c$ with respect to $T$, increases with $d$, in accord with recent simulations for soft-sphere liquids for $d=2$, $3$, and $4$~\cite{JCP_138_12A548}. Analyses of the isothermal compressibility and thermal expansion coefficient, taken to be physical metrics of packing frustration, indicate that molecular packing becomes more efficient in higher $d$, and, as anticipated, these thermodynamic properties strongly correlate with the $d$ dependence of the fragility. We thus confirm the qualitative \textit{geometrical} origin of the reduced kinetic fragility upon increasing $d$.

We consider a $d$-dimensional melt of chains with the structure of polypropylene (PP)~\cite{ACP_137_125}. For comparative purposes, a common parameter set is used in all $d$: the microscopic cohesive energy is $\epsilon=200$ K, the bending energy is $E_b=800$ K, and the polymerization index is $N_c=8000$. By employing these parameters, the GET calculations for constant pressure ($P=0$ MPa) polymer melts for $d=3$ yield the glass transition temperature $T_g$ and fragility~\footnote{The GET defines $T_g$ by $\tau_{\alpha}(T_g)=100$ s, and the isobaric fragility parameter follows as $m_P=\left.\partial \log (\tau_{\alpha})/\partial (T_g/T)\right|_{P, T=T_g}$. Using the parameters in the present work leads to $T_g=313.5$ K and $m_P=148.5$ in $d=3$.} which are typical of synthetic GF polymers.

The temperature dependence of the configurational entropy is central to the AG model. Hence, we first focus on this quantity in $d$ dimensions. The GET considers the configurational entropy density $s_c$ (i.e., the configurational entropy per lattice site~\cite{ACP_137_125}) at constant $P$,
\begin{eqnarray}
	P=-\left.\frac{\partial F}{\partial V}\right|_{N_p,T}=-\left.\frac{1}{V_{\text{cell}}}\frac{\partial F}{\partial N_l}\right|_{N_p,T},
\end{eqnarray}
where $F$ is the Helmholtz free energy~\cite{ACP_103_335}, $V=N_lV_{\text{cell}}$ is the total volume of the system with $N_l$ and $V_{\text{cell}}$ the number of total lattice sites and the volume associated with a single lattice site, and $N_p$ is the number of polymer chains. While $V_{\text{cell}}$ in $d$ dimensions may be defined in terms of the volume $V_{\text{cell}}=a_{\text{cell}}^d$ of a $d$-dimensional cube with $a_{\text{cell}}$ the edge length, basic quantities, such as Boltzmann's constant $k_B$, cannot be determined in $d>3$. We surmount this problem by performing all calculations at $P=0$ MPa in all $d$ since this condition completely eliminates problems in estimating $V_{\text{cell}}$ and $k_B$ in $d>3$.

\begin{figure}[tb]
	\centering
	\includegraphics[angle=0,width=0.45\textwidth]{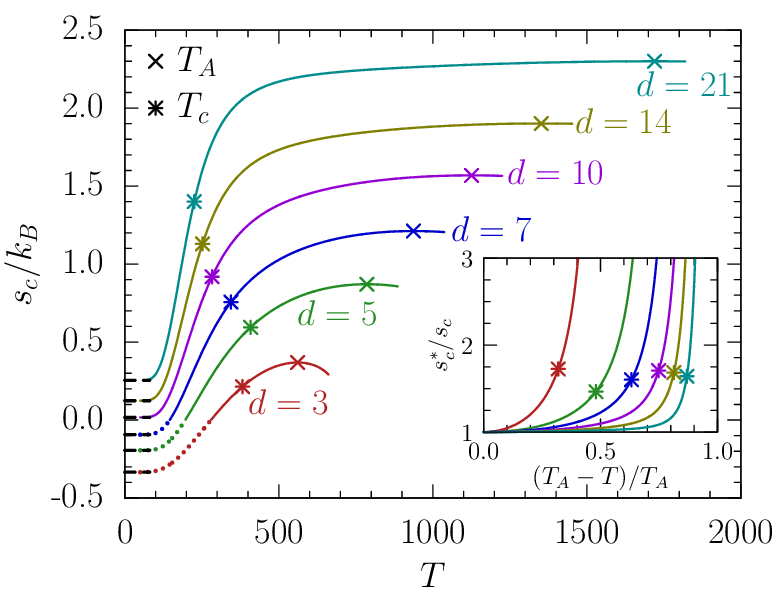}
	\caption{Main: $s_c/k_B$ as a function of $T$ for various $d$ at $P=0$ MPa. Dotted lines highlight where $s_c$ is negative, an unphysical behavior that is probably an artifact of high $T$ expansion in the LCT for low $d$ and a mathematical feature that disappears for higher $d$. Dashed lines denote the low $T$ constants of  $s_{c}/k_B$ for each $d$. Inset: $s_c^*/s_c$ as a function of $(T_A-T)/T_A$ for various $d$.}
\end{figure}

Figure 1 displays the calculated $s_c/k_B$ as a function of $T$ for various $d$. As in $d=3$, $s_c$ for larger $d$ displays a shallow maximum $s_c^\ast$ (crosses in Fig. 1) at a high $T$ that defines the onset temperature $T_A$ of glass-formation in the GET that signals the onset of non-Arrhenius behavior of $\tau_{\alpha}$~\cite{ACP_137_125}, and the product $s_cT$ exhibits an inflection point (asterisks in Fig. 1) at a lower $T$ that uniquely defines the crossover temperature $T_c$ in the GET that separates two regimes of $T$ with qualitatively different dependences of $\tau_{\alpha}$ on $T$~\cite{ACP_137_125}. Evidently, $s_c$ formally becomes negative (dotted lines in Fig. 1) at low $T$ and low $d$ in our mean-field theory, an unphysical behavior that is probably an artifact of high $T$ expansion implicit to the LCT~\cite{JCP_141_234903}. However, taken at face value, the point at which $s_c$ vanishes signals a divergence of $\tau_{\alpha}$ in the AG model. The existence of an equilibrium solidification transition in a disordered material implies that the transition must have the mathematical character of a second-order phase transition, exactly the interpretation made long ago for the glass transition by Gibbs and DiMarzio~\cite{JCP_28_373}.

The GET generally predicts the presence of a low $T$ plateau in $s_c$ (denoted by $s_{c,r}$; see the dashed lines in Fig. 1). This plateau increases linearly with $\ln(d/2)$ and becomes \textit{positive} beyond a critical dimension $d_c$~\cite{SM}, and hence, an ``ideal glass transition temperature'' $T_0$, where $s_c$ formally vanishes, only exists for $d<d_c$ in the GET. We further find that $d_c$ is independent of $\epsilon$ and $E_b$ but varies with $N_c$ for a fixed molecular structure and that $d_c$ is near $d=8$ for the PP structure~\cite{SM}. Some recent models of glass-formation~\cite{JPCM_19_205101, EPJE_34_102} predict a critical dimension of $d_c=8$ above which the Stokes-Einstein relation holds, a prediction that might be related to the behavior displayed in Fig. 1.

The evidence for a nonzero residual entropy in the glassy state has been a matter of long-standing discussion~\cite{JCP_132_124509}. According to the AG theory, the appearance of a positive constant $s_c$ implies that $\tau_{\alpha}$ does not diverge at $T>0$ K and that structural relaxation becomes Arrhenius again at low $T$, albeit with a much higher activation free energy than that above $T_A$. The ratio of the Arrhenius activation free energies in the low and high $T$ Arrhenius regimes above $d_c$ can be calculated as $s_c^*/s_{c,r}$, a quantity that is evidently larger than unity and that decreases with $d$ as $s_c^*/s_{c,r}=1/[A-B/\ln(d/2)]$ with the fitted constants, $A=0.325$ and $B=0.505$~\cite{SM}. Since $\tau_{\alpha}$ becomes astronomically large in the low $T$ Arrhenius regime, the thermodynamic state in this regime can be considered to be a ``glass''. Notably, $s_c$ appears to exhibit a step drop towards $s_{c,r}$ near $T_c$ for large $d$, suggesting that the fluid transforms into a glass at a finite $T$ upon cooling through a first-order phase transition in the $d\rightarrow\infty$ limit. Interestingly, the ``tiling model'' of glass-formation, developed by Weber and Stillinger~\cite{PRB_34_7641, PRB_36_7043}, predicts the glass transition to be first-order. 

We now consider the $T$ dependence of $\tau_{\alpha}$ in $d$ dimensions, which is obtained in the GET through the AG relation~\cite{JCP_43_139},
\begin{equation}
	\tau_{\alpha}=\tau_\infty\exp[\varepsilon_{\text{AG}}/(k_BT)],~\varepsilon_{\text{AG}}=\Delta\mu (s_c^\ast/s_c),
\end{equation}
where $\tau_\infty$ is the high $T$ limit of $\tau_{\alpha}$, $\varepsilon_{\text{AG}}$ is the AG activation free energy, and $\Delta\mu$ is the $T$ independent activation free energy at high $T$ above $T_A$ where the Arrhenius dependence of $\tau_{\alpha}$ holds. $\tau_\infty$ in $d=3$ is taken in the GET as a ``default'' value of $10^{-13}$ s. Since $\tau_\infty$ is expected to be non-universal in variable $d$, we consider only the dimensionless relaxation time $\tau_{r}=\tau_{\alpha}/\tau_\infty$. The GET estimates $\Delta\mu$ in $d=3$ from the empirical relation, $\Delta\mu/k_B=6~T_c$~\cite{ACP_137_125}. Simulations for $d=2$, $3$, and $4$~\cite{JCP_138_12A548, PRE_85_061501} suggest that $\Delta\mu$ increases approximately in proportion to $d$. This phenomenology is quite understandable from classical arguments by Eyring~\cite{CR_28_301} relating $\Delta\mu$ to the cohesive interaction strength of the fluid, which scales in a hypercubic lattice model in proportion to the coordination number ($2d$). The scaling of $\Delta\mu$ with $2d$ implies that $\Delta\mu$ in the GET obeys $\Delta\mu/k_B=2d~T_{c,d=3}$, where $T_{c,d=3}$ is the $T_c$ in $d=3$.

Before analyzing the $T$ dependence of $\tau_{r}$, we investigate the $T$ dependence of $s_c^*/s_c$ in $d$ dimensions, which describes the multiplicative increase of $\varepsilon_{\text{AG}}$. The inset to Fig. 1 displays $s_c^*/s_c$ for various $d$ as a function of $(T_A-T)/T_A$, a reduced temperature that measures the degree of quench into the ``conjested'' regime of glassy dynamics. The growth of $\varepsilon_{\text{AG}}$ with increasing $d$ and, correspondingly with the scale of collective motion, weakens for fixed $(T_A-T)/T_A$. Evidently, the GET predicts that structural relaxation in the fluid state becomes Arrhenius in the $d\rightarrow\infty$ limit where $s_c^*/s_c$ approaches unity asymptotically. Previous work~\cite{ACP_137_125} indicates that $s_c^*/s_c$ for $d=3$ exhibits a parabolic dependence on $(T_A-T)/T_A$, 
\begin{equation}
s_c^*/s_c=1+C_s[(T_A-T)/T_A]^2,~T_c<T<T_A.
\end{equation}
$C_s$ measures the extent of cooperative motion at high $T$ in the GET~\cite{ACP_137_125}, and we find that Eq. (3) holds well in ($T_A-50$, $T_A$) in $d\ge 3$.

\begin{figure}[tb]
	\centering
	\includegraphics[angle=0,width=0.45\textwidth]{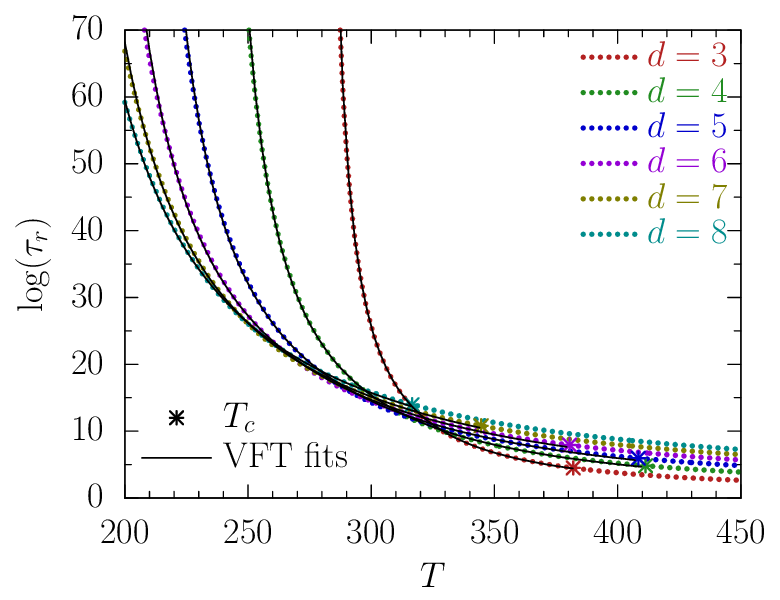}
	\caption{$\log(\tau_{r})$ as a function of $T$ for various $d$.}
\end{figure}

\begin{figure}[b]
	\centering
	\includegraphics[angle=0,width=0.45\textwidth]{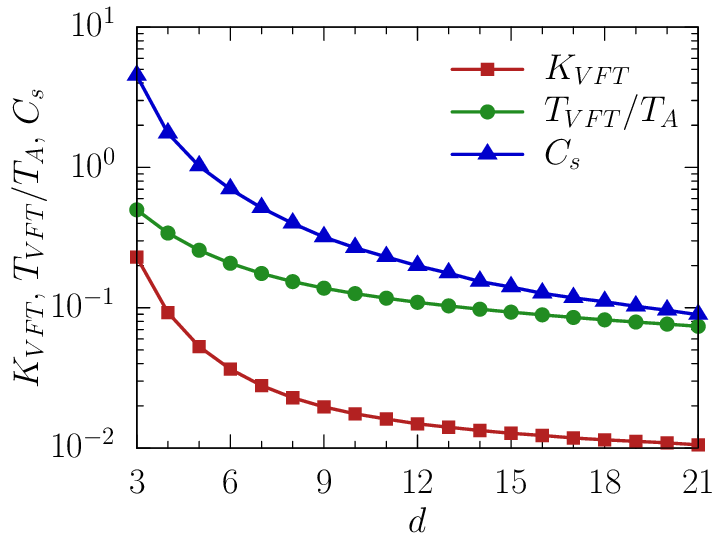}
	\caption{$K_{\text{VFT}}$, $T_{\text{VFT}}/T_A$, and $C_s$ as a function of $d$.}
\end{figure}

Figure 2 displays the $T$ dependence of $\log(\tau_{r})$ for various $d$, along with the fits to the Vogel-Fulcher-Tammann (VFT) equation,
\begin{equation}
	\tau_{r}=\tau_0\exp\left[\frac{1}{K_{\text{VFT}}(T/T_{\text{VFT}}-1)}\right],
\end{equation}
where $\tau_0$, $K_{\text{VFT}}$, and $T_{\text{VFT}}$ designate the high $T$ limit of $\tau_{r}$, the ``kinetic fragility'' parameter quantifying the degree to which relaxation is non-Arrhenius, and the VFT temperature where $\tau_{r}$ extrapolates to infinity, respectively. Previous analyses for $d=3$~\cite{ACP_137_125} have shown that the VFT equation describes $\tau_{\alpha}$ calculated from the GET quite well within the $T$ range between $T_c$ and $T_g$. Since we are unable to define $T_g$ in $d>3$ in terms of a characteristic timescale, VFT fits are performed for the $T$ range between $T_c$ and a lower temperature $T_l$ at which $\log[\tau_{r}(T_l)]=63$~\footnote{$T_l$ corresponds to a large $\tau_{r}$ since $\tau_{r}$ at $T_c$ is already large for high $d$. We have confirmed that the general trend of the fitted $K_{\text{VFT}}$ and $T_{\text{VFT}}$ with $d$ is unaffected by the reasonable choice of $T_l$.}. The $d$ dependence of $K_{\text{VFT}}$ is depicted in Fig. 3, along with $C_s$ and the ratio $T_{\text{VFT}}/T_A$, a measure of the ``breadth'' of glass-formation~\cite{ACP_137_125}. All these three metrics for glassy dynamics {\it{decrease}} with $d$. 

\begin{figure}[tb]
	\centering
	\includegraphics[angle=0,width=0.45\textwidth]{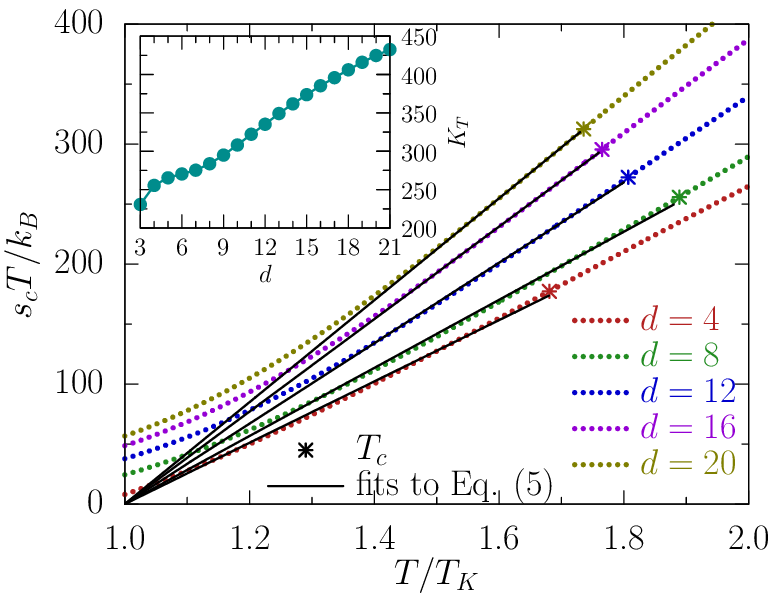}
	\caption{Main: $s_cT/k_B$ as a function of $T/T_K$ for various $d$. Inset: $K_T$ as a function of $d$.}
\end{figure}

Following the simulation studies~\cite{JCP_138_12A548}, we additionally explore how $d$ alters the ``thermodynamic fragility'' parameter $K_T$, determined from the slope of the $T$ dependence of the product $s_cT$,
\begin{equation}
s_cT/k_B=K_T(T/T_K-1),
\end{equation}
where $T_K$ is the Kauzmann temperature, estimated by linearly extrapolating $s_cT$ to zero over the same $T$ range where $K_{\text{VFT}}$ is obtained (see Fig. 4). We prefer to call $K_T$ the ``low $T$ thermodynamic steepness'' parameter because this parameter bears no direct relation to the strength of the $T$ dependence of $\tau_{\alpha}$. The inset to Fig. 4 indicates that $K_T$ \textit{increases} with $d$, in agreement with the trend found in recent simulations of soft-sphere GF liquids in $d=2$, $3$, and $4$~\cite{JCP_138_12A548}. Reference~\cite{JCP_138_12A548} also shows, based on limited data from simulations, that $K_{\text{VFT}}$ seems to weakly increase with $d$ in the same model~\footnote{Unfortunately, $K_{\text{VFT}}$ is estimated in simulations using data above $T_c$~\cite{JCP_138_12A548}, while $K_{\text{VFT}}$ is obtained in our model from data below $T_c$. Moreover, simulations~\cite{JCP_138_12A548} and our calculations are performed at constant density and constant pressure, respectively. Hence, it is difficult to directly compare $K_{\text{VFT}}$ in our model and simulations.}. 

Our theory, thus, predicts that $K_{\text{VFT}}$ varies in the \textit{opposite} direction as $K_T$ when $d$ grows. An opposite dependence of $K_{\text{VFT}}$ and $K_T$ is also found in $3$-dimensional simulations of GF liquids in which particles interact with a modified Lennard-Jones potential~\cite{JCP_135_194503}, where the potential softness is tuned to alter the fragility and where the inverted variations of $K_T$ and $K_{\text{VFT}}$ have been attributed to variations in $\Delta\mu$. Apparently, the same mechanism also explains our results since $\Delta\mu$ scales with $2d$ so that $\Delta\mu$ can be the dominant contribution to $K_{\text{VFT}}$ for high $d$.

\begin{figure}[tb]
	\centering
	\includegraphics[angle=0,width=0.48\textwidth]{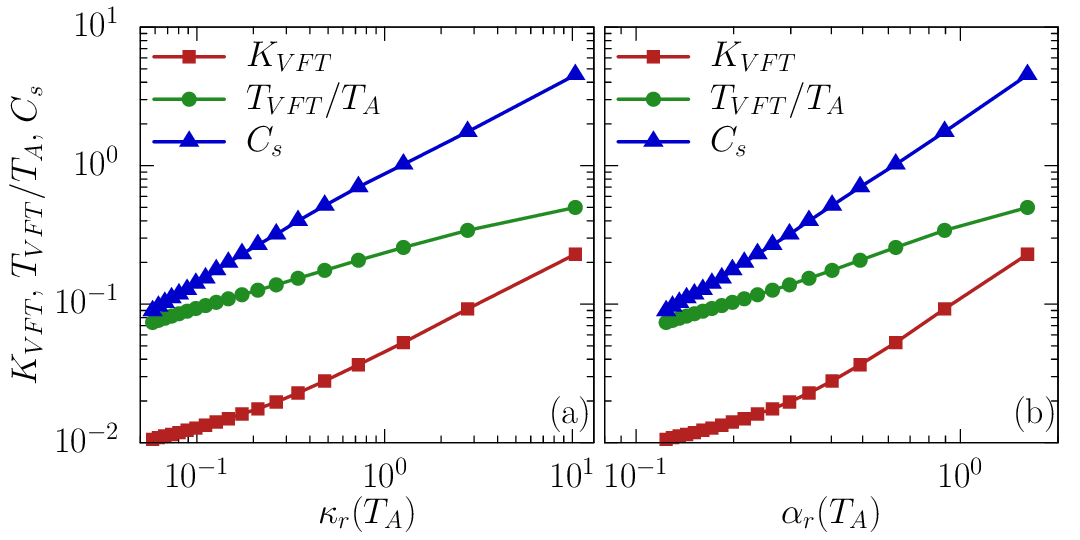}
	\caption{Correlations between $K_{\text{VFT}}$, $T_{\text{VFT}}/T_A$, $C_s$ and (a) $\kappa_r(T_A)$ and (b) $\alpha_r(T_A)$.}
\end{figure}

We now seek what physical factors govern the fragility of glass-formation. Building on the earlier GET calculations for $d=3$ indicating that the fragility of polymer GF liquids primarily arises from packing frustration~\cite{ACP_137_125}, we explore physical metrics that might best describe packing frustration. We consider the isothermal compressibility $\kappa_T=-(1/V)\left. (\partial V/\partial P)\right|_T$ and the thermal expansion coefficient $\alpha_P=(1/V)\left. (\partial V/\partial T)\right|_P$ as possible physical measures of packing frustration. In particular, we analyze $\kappa_r=\kappa_T(k_B T)/\rho$ (where $\rho=\varphi~V_{\text{cell}}$ with $\varphi$ the polymer filling fraction~\cite{ACP_103_335}) and $\alpha_r=T\alpha_P$, since these reduced quantities are found to strongly correlate with fragility in $d=3$. (A more detailed discussion for $d=3$ will be presented elsewhere.) As expected, as $d$ grows, the $T$ dependence of $\kappa_r$ and $\alpha_r$ weakens and their overall magnitudes greatly decreases, thereby reflecting the diminished packing frustration with increasing $d$~\cite{SM}. As an illustration, Fig. 5 exhibits the presence of strong correlations between various dynamic metrics ($K_{\text{VFT}}$, $T_{\text{VFT}}/T_A$, and $C_s$) and $\kappa_r(T_A)$ and $\alpha_r(T_A)$, implying that the $d$ dependence of fragility can be predicted from a knowledge of the physical measures of packing frustration. Therefore, both the reduced fragility with increasing $d$ and the corresponding return to Arrhenius relaxation are consequences of the reduction of packing frustration in higher $d$ where space is more ``open'' and packing constraints are more weakly ``felt''.

In summary, our exploration of polymer glass-formation in variable dimension indicates some clear trends as the dimensionality is varied. Perhaps most basically, we find that structural relaxation in the fluid state asymptotically becomes Arrhenius in the $d\rightarrow\infty$ limit and that the fluid transforms upon cooling above a critical dimension near $d=8$ into an amorphous state with a finite positive residual configurational entropy. The transition from a fragile glass-forming liquid in low dimensions to an ideally strong liquid in the $d\rightarrow\infty$ limit indicates that the kinetic fragility decreases with increasing dimension. In accord with simulations~\cite{JCP_138_12A548}, our theory predicts that the thermodynamic fragility $K_T$ grows at elevated $d$. $K_T$ fails to be a fragility parameter, and we suggest that $K_T$ be termed the ``low temperature thermodynamic steepness'' parameter. By establishing strong correlations between the isothermal compressibility and thermal expansion coefficient and the fragility, we suggest that these thermodynamic properties should provide physical metrics of packing frustration for predicting fragility variations in glass-forming liquids. 

\begin{acknowledgments}
We thank Jacek Dudowicz for his numerous contributions to the development of the entropy theory of glass-formation that have helped make the present work possible. This work is supported by the U.S. Department of Energy, Office of Basic Energy Sciences, Division of Materials Sciences and Engineering under Award DE-SC0008631.
\end{acknowledgments}


\bibliography{refs}

\pagebreak
\clearpage
\begin{center}
	\textbf{\large Supplemental Material for\\
		Polymer Glass-Formation in Variable Dimension}
\end{center}
\setcounter{equation}{0}
\setcounter{figure}{0}
\setcounter{table}{0}
\setcounter{page}{1}
\setcounter{section}{0}
\makeatletter
\renewcommand{\thepage}{S\arabic{page}}  
\renewcommand{\thesection}{S\arabic{section}}   
\renewcommand{\thetable}{S\arabic{table}}   
\renewcommand{\thefigure}{S\arabic{figure}}
\renewcommand{\theequation}{S\arabic{equation}}

\section{Mean-field estimation of the configurational entropy at low temperatures and dependence of the critical dimensionality on molar mass}

The residual configurational entropy density $s_{c,r}$ [i.e., the low temperature ($T$) constant of the configurational entropy density $s_{c}$] can be obtained by analyzing the thermodynamics of the system in the limit where all lattice sites are occupied by chain segments. Hence, we focus on the low $T$ regime of glass-formation where the polymer filling fraction $\varphi$ (defined as the ratio of the total number of chain segments to the total number of lattice sites~\cite{ACP_103_335, JCP_141_044909}) approaches unity in order to estimate how the critical dimensionality, above which $s_c$ ceases to vanish, depends on molecular parameters. For low $T$ (e.g., $T=10$ K), the factor $\exp(-\beta E_b)$ [where $\beta=1/(k_BT)$ and $E_b$ is the bending energy] in the LCT free energy nearly vanishes for large $E_b$ [e.g., the present work uses $E_b=800$ K, leading to $\exp(-\beta E_b)=\exp(-80)$ at $T=10$ K]. In this limit, all terms multiplied by $\exp(-\beta E_b)$ in the LCT free energy disappear, and then taking $\varphi=1$ leads to the greatly simplified expression for the free energy,
\begin{equation}
	\beta f_r=\frac{1}{M}\ln\left(\frac{2}{z^{L}M}\right)+\left(1-\frac{1}{M}\right)-\sum_{i=1}^6C_i,
\end{equation}
where $M$ is the molar mass (defined as the total number of chain segments in a single chain), $L$ designates the number of sub-chains in a single chain~\cite{ACP_103_335, JCP_141_044909}, and $C_i$ has the following form~\cite{JCP_141_044909}:
\begin{equation}
	C_i=C_{i, 0}+C_{i, \epsilon}(\beta\epsilon)+C_{i, \epsilon^2}(\beta\epsilon)^2,
\end{equation}
where $\epsilon$ is the nearest-neighbor van der Waals interaction energy. However, $C_{i, 0}$, $C_{i, \epsilon}$, and $C_{i, \epsilon^2}$ are now independent of $T$, and, hence, 
\begin{equation}
	\beta \frac{\partial C_i}{\partial \beta}=C_{i, \epsilon}(\beta\epsilon)+2C_{i, \epsilon^2}(\beta\epsilon)^2.
\end{equation}
Consequently, the residual configurational entropy density, $s_{c,r}=-\left. \partial f_r/\partial T\right|_{\varphi}$, emerges as
\begin{eqnarray}
	\frac{s_{c,r}}{k_B}=&&-\frac{1}{M}\ln\left(\frac{2}{z^{L}M}\right)-\left(1-\frac{1}{M}\right)\nonumber\\
	&&
	+\sum_{i=1}^6\left(C_i-\beta \frac{\partial C_i}{\partial \beta}\right).
\end{eqnarray}
Substituting Eq. (S3) into Eq. (S4) leads to the relation,
\begin{eqnarray}
	\frac{s_{c,r}}{k_B}=&&-\frac{1}{M}\ln\left(\frac{2}{z^{L}M}\right)-\left(1-\frac{1}{M}\right)\nonumber\\
	&&
	+\sum_{i=1}^6\left[C_{i, 0}-C_{i, \epsilon^2}(\beta\epsilon)^2\right].
\end{eqnarray}
Taking advantage of the fact that $\sum_{i=1}^6C_{i, \epsilon^2}=0$~\cite{JCP_141_044909}, $s_{c,r}$ finally appears in the simple form:
\begin{eqnarray}
	\frac{s_{c,r}}{k_B}=-\frac{1}{M}\ln\left(\frac{2}{z^{L}M}\right)-\left(1-\frac{1}{M}\right)+\sum_{i=1}^6C_{i, 0}.
\end{eqnarray}

\begin{figure}[tb]
	\centering
	\includegraphics[angle=0,width=0.45\textwidth]{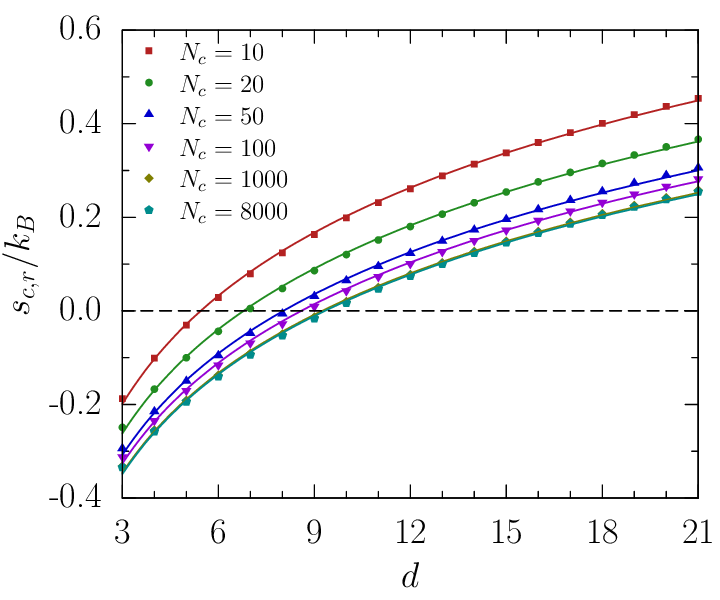}
	\caption{Residual configurational entropy density $s_{c,r}/k_B$ as a function of $d$ for various polymerization indices $N_c$ for a melt of PP chains. The solid lines are fits to the equation $s_r/k_B=A\ln(d/2)+B$ with $A$ and $B$ being the fitting parameters. The critical dimensionality $d_c$, where $s_{c,r}$ starts to be positive, increases with $N_c$ and saturates at $d=10$ for sufficiently large $N_c$.}
\end{figure}

Since $C_{i, 0}$ is now a function of $d$ ($z=2d$) and a set of geometrical indices that are determined from the monomer structure and molar mass $M$, $s_{c,r}$ is independent of $T$, $\epsilon$, and $E_b$. Figure S1 displays the $d$ dependence of $s_{c,r}$ for various polymerization indices $N_c$ for a melt of PP chains. The ``critical dimensionality'' $d_c$, where $s_{c,r}$ starts to be positive, is found to increase with $N_c$, saturating at $d=10$ for large $N_c$. The $d$ dependence of $s_{c,r}/k_B$ in each case can be well described by the equation,
\begin{eqnarray}
	s_{c,r}/k_B=A\ln(d/2)+B,
\end{eqnarray}
where $A$ and $B$ are the fitting parameters. Figure S1 illustrates this variation. The filling fraction at the point at which $s_c$ formally vanishes is somewhat less than unity, but this simple analytic argument is sufficient for estimating the $N_c$ dependence of the critical dimensionality above which $s_c$ no longer vanishes at low $T$.

\section{Dimensional dependence of the ratio of the activation free energies in the high and low temperature Arrhenius regimes}

\begin{figure}[tb]
	\centering
	\includegraphics[angle=0,width=0.45\textwidth]{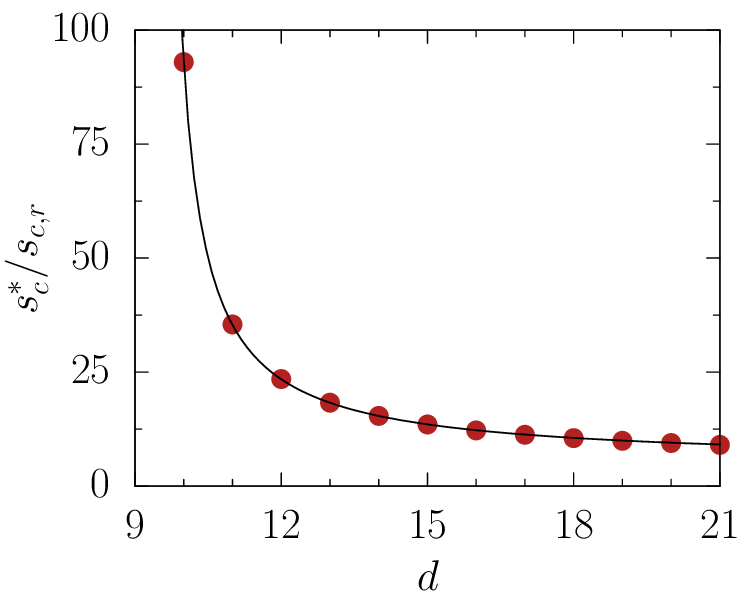}
	\caption{Activation free energy ratio $s_c^*/s_{c,r}$ as a function of $d$ for $d\ge d_c$. The line is a fit to the equation $s_c^*/s_{c,r}=1/[A-B/\ln(d/2)]$ with the fitted constants, $A=0.325$ and $B=0.505$.}
\end{figure}

As discussed in the main text, the ratio $s_c^*/s_{c,r}$ defines the relative magnitude of the activation free energy for relaxation in the high and low $T$ Arrhenius regimes above $d_c$ that we separately characterize as being simple fluid-like and glass-like because of the large difference of relaxation times in these different regimes of $T$. The GET allows us to calculate this activation free energy ratio as a function of $d$, and we illustrate our findings in Fig. S2. Curiously, the activation free energy ratio falls with increasing $d$, indicating that the dynamics of the high and low $T$ regimes actually becomes similar in high $d$, where the activation free energy ratio saturates at a minimal value of about $3$. The $d$ variation of the activation free energy ratio is well represented by the approximant:
\begin{eqnarray}
	s_c^*/s_{c,r}=1/[A-B/\ln(d/2)],~d>d_c,
\end{eqnarray}
where the fitting parameters are estimated to be $A=0.325$ and $B=0.505$. Because the activation free energy in the high $T$ regime is large for high $d$, increasing this quantity by a factor of $3$ in high $d$ translates into an extremely large increase in relaxation times in the low $T$ Arrhenius regime.

\section{Dimensional dependence of the reduced isothermal compressibility and thermal expansion coefficient}

\begin{figure}[htb]
	\centering
	\includegraphics[angle=0,width=0.45\textwidth]{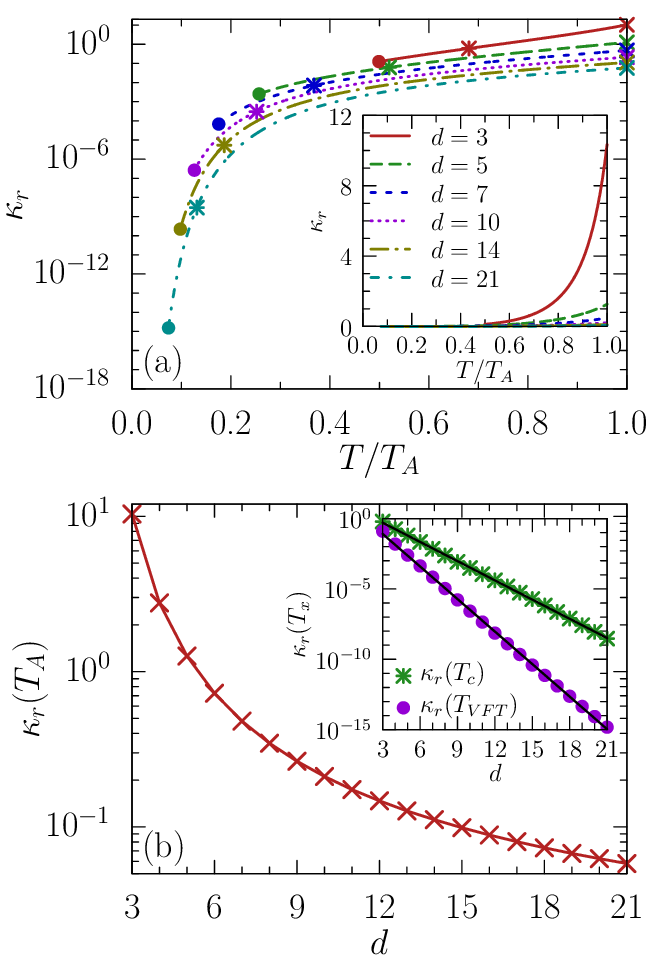}
	\caption{(a) Main: Reduced isothermal compressibility $\kappa_r=\kappa_T(k_B T)/\rho$ as a function of $T/T_A$ for various $d$. Crosses, asterisks, and circles indicate the points corresponding to $T_A$, $T_c$, and $T_{\text{VFT}}$, respectively. Data are presented in the $T$ range ($T_{\text{VFT}}$, $T_A$) of interest in the present work. Inset: Same plot but shown in a linear scale representation. (b) Main: $\kappa_r(T_A)$ as a function of $d$. Inset: $\kappa_r(T_c)$ and $\kappa_r(T_{\text{VFT}})$ as a function of $d$. Lines in the set indicate the exponential fits, $\kappa_r(T_{x})=A_{x}\exp(-B_{x} d)$ ($x=c$ or $\text{VFT}$) with $A_{x}$ and $B_{x}$ the fitting parameters. We obtained the fitting estimates: $A_{c}=11.7$ and $B_{c}=1.05$ and $A_{\text{VFT}}=16.3$ and $B_{\text{VFT}}=1.77$.}
\end{figure}

\begin{figure}[htb]
	\centering
	\includegraphics[angle=0,width=0.45\textwidth]{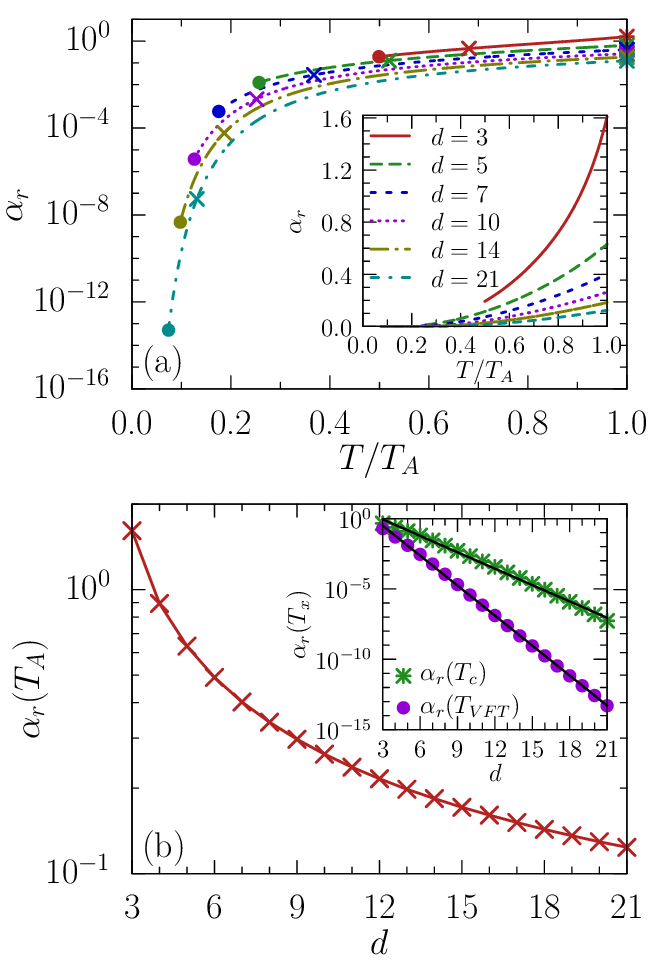}
	\caption{(a) Main: Reduced thermal expansion coefficient $\alpha_r=T\alpha_P$ as a function of $T/T_A$ for various $d$. Crosses, asterisks, and circles indicate the points corresponding to $T_A$, $T_c$, and $T_{\text{VFT}}$, respectively. Data are presented in the $T$ range ($T_{\text{VFT}}$, $T_A$) of interest in the present work. Inset: Same plot but shown in a linear scale representation. (b) Main: $\alpha_r(T_A)$ as a function of $d$. Inset: $\alpha_r(T_c)$ and $\alpha_r(T_{\text{VFT}})$ as a function of $d$. Lines in the set indicate exponential fits: $\alpha_r(T_{x})=A_{x}\exp(-B_{x} d)$ ($x=c$ or $\text{VFT}$) with $A_{x}$ and $B_{x}$ the fitting parameters. We obtained the fitting estimates: $A_{c}=12.7$ and $B_{c}=0.89$ and $A_{\text{VFT}}=42.1$ and $B_{\text{VFT}}=1.63$.}
\end{figure}

Materials that pack more efficiently can be expected to have a lower and more weakly $T$ dependent isothermal compressibility ($\kappa_T$) or thermal expansion coefficient ($\alpha_P$), while those exhibiting a high degree of packing frustration due to the complexity of the molecular structure should exhibit a relatively large and relatively strongly $T$ dependent $\kappa_T$ or $\alpha_P$. Previous analyses~\cite{JCP_131_114905} of GET calculations in $d=3$ reveal the existence of a very strong correlation between the filling fraction at $T_g$ and the fragility when the molar mass, cohesive interaction, or pressure is varied. Moreover, molecules in fragile polymer fluids are predicted to pack less efficiently in the glassy state than those in stronger glass-forming polymer fluids. These observations deduced from the GET calculations indicate that packing efficiency is a central factor for understanding the fragility of glass-formation. This line of reasoning naturally leads to the expectation that $\kappa_T$ and $\alpha_P$ correlate strongly with the fragility.

Because our extensive analyses in $d=3$ reveal that the dimensionless isothermal compressibility, $\kappa_r=\kappa_T(k_B T)/\rho$ with $\rho=\varphi~V_{\text{cell}}$, and the dimensionless thermal expansion coefficient, $\alpha_r=T\alpha_P$, strongly correlate with the fragility of glass-formation, we consider these reduced quantities in $d$ dimensions. Figures S3 and S4 indicate the GET model predictions for $\kappa_r$ and $\alpha_r$ as a function of $T/T_A$ for various $d$. Evidently, the strength of the $T$ dependence of both $\kappa_r$ and $\alpha_r$ weakens with increasing $d$, and both $\kappa_r$ and $\alpha_r$ progressively decreases with $d$ for fixed reduced temperatures $T/T_A$. The insets to Figs. S3 and S4 indicate that $\kappa_r(T_x)$ ($x=c$ or $\text{VFT}$) and $\alpha_r(T_x)$ ($x=c$ or $\text{VFT}$) decay exponentially with $d$ to a high approximation. The slower variation with $T$ of $\kappa_r$ and $\alpha_r$ and the overall values of $\kappa_r$ and $\alpha_r$ all reflect decreased packing frustration with increasing $d$, thereby confirming the qualitative \textit{geometrical} origin of the reduced fragility found upon increasing $d$. Both the reduced fragility with increasing $d$ and the corresponding return to Arrhenius relaxation are consequences of the reduction of packing frustration in high $d$ where space is more ``open'' and packing constraints are more weakly ``felt''.

\end{document}